\documentstyle[preprint,aps]{revtex}
\tighten
\draft
\widetext
\input epsf
\preprint{HUTP-97/A015, NUB 3157}
\bigskip
\bigskip
\begin{document}
\title{Aspects of $N=1$ Type I-Heterotic Duality in Four Dimensions}
\medskip
\author{Zurab Kakushadze\footnote{E-mail: zurab@string.harvard.edu}}
\bigskip
\address{Lyman Laboratory of Physics, Harvard University, Cambridge, MA 02138\\
and\\
Department of Physics, Northeastern University, Boston, MA 02115}
\date{April 7, 1997}
\bigskip
\medskip
\maketitle

\begin{abstract}
{}In this paper we discuss some aspects of $N=1$ type I-heterotic string duality in four dimensions. We consider a particular example of a (weak-weak) dual pair where on the type I side there are only $D9$-branes corresponding to perturbative heterotic description in a certain region of the moduli space. We match the perturbative type I and heterotic tree-level massless spectra via giving certain scalars appropriate vevs, and point out the crucial role of the perturbative superpotential (on the heterotic side) for this matching. We also discuss the role of anomalous $U(1)$ gauge symmetry present in both type I and heterotic models. In the perturbative regime we match the (tree-level) moduli spaces of these models. Since both type I and heterotic models can be treated perturbatively, we are able to discuss a dictionary that in generic models maps type I description onto heterotic one, and vice-versa. Finally, we discuss possible directions to study perturbative quantum corrections to the moduli space, as well as outline ways to learn about the non-perturbative effects in both descriptions.    
\end{abstract}

\section{Introduction}

{} Non-perturbative string effects are very likely to be crucial for (re)formulating string theory in a way that might be suitable for describing the nature. In the recent years it has become clear that understanding non-perturbative string dynamics may be within our reach. String dualities have been playing a key role in making progress along these lines, as they allow one to cast (at least in certain cases) non-perturbative effects in one theory into perturbative ones in its dual, and vice-versa. We have learned a great deal about $N=4$ and $N=2$ string dualities (in four dimensions), and some progress has also been made in trying to understand $N=1$ cases. The latter are much more involved than the former for the larger the number of unbroken supersymmetries, the more control we have over non-perturbative string dynamics. Yet, one hopes that perhaps $N=1$ string dualities, although almost certainly more complex, might still be within grasp. This hope is also supported by recent developments in $N=1$ field theory dualities \cite{Seiberg}, including their understanding from string theory perspective \cite{Vafa}.

{}One case that we might be able to understand thoroughly is $N=1$ type I-heterotic string duality in four dimensions. There are at least a few reasons to believe that this may be so. Thus, the tree-level relation between type I and heterotic dilatons in $D$ space-time dimensions \cite{Sagnotti} (which follows from the conjectured type I-heterotic duality in ten dimensions
\cite{typeI-het-10}) reads:
\begin{equation}
 \phi_H={{6-D}\over 4} \phi_I -{{D-2}\over 16}\log(\det(g_I))~.
\end{equation} 
Here $g_I$ is the internal metric of the type I compactification space, whereas
$\phi_I$ and $\phi_H$ are the type I and heterotic dilatons, respectively. From this one can see that (in four dimensions) there always exists a region in the moduli space where both type I and heterotic string theories are weekly coupled, and there we can rely on perturbation theory, which we understand. If we manage to identify the relation between the perturbative effects in the two descriptions, we may hope to be able to learn about non-perturbative effects in one of the theories, say, heterotic string, by studying the cases where these effects are perturbative on the type I side. An example of this would be (non-perturbative) dynamics of five-branes in heterotic string that can presumably be understood by studying (perturbative) dynamics of type I $D5$-branes.

{}With this goal in mind, in this paper we attempt to understand certain aspects of $N=1$ type I-heterotic string duality in four dimensions. We will do this on a particular example, namely, the chiral $N=1$ type I model constructed in Ref \cite{Sagnotti}, and its (candidate) heterotic dual that was also proposed in Ref \cite{Sagnotti}. One of the issues that need to be resolved for
making the duality between these two models precise is to match their (tree-level) massless spectra, which are {\em not} the same at the points of the moduli space where the original models of Ref \cite{Sagnotti} are constructed. In particular, the heterotic model has extra twisted matter fields that do not seem to have type I counterparts. We show that the perturbative tree-level superpotential present in the heterotic model plays a crucial role in giving masses to these extra fields after appropriate Higgsing. This feature seems to be quite general for $N=1$ string dualities. In particular, perturbative superpotentials seem to be necessary to match the massless spectra of $F$-theory and (candidate) heterotic dual pairs \cite{Bershadsky}. It is only natural that this is the case for $N=1$ models where perturbative superpotentials are generically present. This is to be contrasted with cases of larger space-time suppersymmetric theories where the dynamics is less complex (at least perturbatively). 

{}We also point out certain subtleties appearing in the type I and heterotic models that we consider in this paper due to the presence of anomalous $U(1)$.
The latter seems to be a generic feature in a certain class of $N=1$ models. Certainly, our long experience with heterotic string models is indicative of this. We believe that anomalous $U(1)$ might play a crucial role for $F$-theory compactifications as well, and it would be interesting to investigate issues associated with its presence, for example, in the context of $N=1$ field theory dualities \cite{Vafa} arising from $F$-theory compactifications on Calabi-Yau four-folds \cite{Bershadsky}.

{}Our choice of models in this paper is motivated by the discussion given in the beginnig of this section. Before plunging into debris of non-perturbative dynamics it is important to understand the dictionary between type I and heterotic perturbative effects. The type I model that we consider has only $D9$-branes (which correspond to perturbative heterotic states) but no $D5$-branes (which would describe non-perturbative heterotic effects due to small instantons). We give such a dictionary in this paper. Our analysis is restricted to the tree-level considerations, and it would also be important to understand perturbative loop corrections, in particular, to the K{\"a}hler potential. At tree-level we do establish equivalence of the type I and heterotic moduli spaces for these models in certain perturbative regime.     

{}The paper is organized as follows. In section II we briefly review the type I model constructed in Ref \cite{Sagnotti}, which is followed by discussion of its candidate heterotic dual in section III. In section IV we give the perturbative superpotentials for these models. In section V we discuss the moduli space, and explain the matching between the type I and heterotic moduli spaces, as well as their tree-level spectra. Section VI is devoted to ``discrete moduli'' of type I theory and its heterotic counterparts. In section VII we give the dictionary between perturbative effects in type I and heterotic strings.
Section VIII contains conclusions and remarks.

\section{Type I Model}

{}In this section we discuss the type I model constructed in Ref \cite{Sagnotti}.
Let us start from the type IIB string model compactified on the six-torus which is
a product of three two-tori, each with ${\bf Z}_3$ rotational symmetry. (A more 
detailed discussion of these two-tori will be given in the next section where we go through the construction of the (candidate) 
heterotic dual of the model considered in this
section.) This model has $N=8$ supersymmetry. Let us now consider the symmetric 
$Z$-orbifold model generated by the twist
\begin{equation}
 T_3=(\theta,\theta,\theta \vert\vert \theta, \theta, \theta)~.
\end{equation} 
Here $\theta$ is a $2\pi /3$ rotation of a complex boson (we have complexified 
two real bosons associated with each of the two-tori). The double vertical line
separates the right- and left-movers of the string. The resulting model has $N=2$ space-time supersymmetry. This model has the following moduli. There are 20 NS-NS fields $\phi,B_{\mu\nu},B_{i{\bar j}},g_{i{\bar j}}$, and 20 R-R fields $\phi^\prime,B^\prime_{\mu\nu},B^\prime_{i{\bar j}},C^\prime_{\mu\nu i{\bar j}}$.

{}Let us now consider the orientifold projection of this model. The closed string sector (which is simply the subspace of the
Hilbert space of the original type IIB spectrum invariant under the orientifold
projection $\Omega$) contains the $N=1$ supergravity multiplet, and 9 untwisted 
(the NS-NS fields that survive the $\Omega$ projection are
$g_{i{\bar j}}$, whereas the R-R fields that are kept are $B^\prime_{i{\bar j}}$; note that the NS-NS field $\phi$ and the R-R field $B^\prime_{\mu\nu}$ also survive, and enter in the dilaton supermultiplet)
and 27 twisted chiral supermultiplets (which are neutral under the gauge group of the model). For consistency (tadpole cancellation) we must include the open string sector. Note that in this model we only have $D9$-branes but no $D5$-branes since the orbifold group does not contain an order two element. (If the orbifold group contains an order two element $R$, then the sector $R\Omega$ would contain $D5$-branes). Thus, we only have $99$ open strings. The gauge group consistent with tadpole cancellation then is $U(12)\otimes SO(8)$ \cite{Sagnotti}. The $99$ open strings also give rise to the following chiral matter fields:
$3({\bf 12}, {\bf 8}_v)(-1)_L$ and $3({\overline {\bf 66}}, {\bf 1})
(+2)_L$. Here the first two entries in bold font indicate the irreps of the
$SU(12)\otimes SO(8)$ subgroup, whereas the $U(1)$ charges  are given in the parenthesis. The factor 
``3'' indicates the number of families (and comes from the excitations of the world-sheet fermions). The subscript $L$ indicates the space-time helicity of the corresponding fermionic fields. The massless spectrum of this model is summarized in Table I.

{}Note that the $U(1)$ gauge symmetry is anomalous. The total $U(1)$ anomaly is $+108$. By the generalized Green-Schwarz
mechanism \cite{GS} some of the fields charged under $U(1)$ will acquire vevs to cancel the Fayet-Illiopoulos $D$-term.

\section{Heterotic String Model}

{}In this section we give the construction of
the heterotic string model that was proposed 
\cite{Sagnotti} to be dual to the type I
model considered in the previous section. Let us start from the Narain model
with $N=4$ space-time supersymmetry in four dimensions. Let the momenta of the
internal (6 right-moving and 22 left-moving) world-sheet bosons span the 
(even self-dual) Narain lattice $\Gamma^{6,22}=\Gamma^{6,6}\otimes\Gamma^{16}$.
Here $\Gamma^{16}$ is the ${\mbox{Spin}}(32)/{\bf Z}_2$ lattice, whereas the 
lattice $\Gamma^{6,6}$ is spanned by the momenta $(p_R \vert\vert p_L)$ with
\begin{eqnarray}\label{momenta}
 p_{L,R}={1\over 2}m_i {\tilde e}^i \pm n^i e_i ~.
\end{eqnarray}
Here $m_i$ and $n^i$ are integers, $e_i \cdot e_j =g_{ij}$ is the constant 
background metric of the compactification manifold (six-torus), and $e_i \cdot
{\tilde e}^j={\delta_i}^j$. Note that we could have included the constant
anti-symmetric background tensor field $B_{ij}$, but for now we will set it equal to
zero for the reasons that will become clear in the following.

{}This Narain model has the gauge group $SO(32) \otimes U(1)^6$. The first 
factor $SO(32)$ comes from the $\Gamma^{16}$ lattice (the $480$ roots of 
length squared 2), and 16 oscillator excitations of the corresponding 
world-sheet bosons (the latter being in the Cartan subalgebra of $SO(32)$).
The factor $U(1)^6$ comes from the oscillator excitations of the six 
left-moving world-sheet bosons corresponding to $\Gamma^{6,6}$. Note that
there are also six additional vector bosons coming from the oscillator 
excitations of the right-moving world-sheet bosons corresponding to 
$\Gamma^{6,6}$. These vector bosons are part of the $N=4$ supergravity 
multiplet.

{}Next consider the $Z$-orbifold model (with non-standard embedding
of the gauge connection) obtained via twisting the above Narain model by the
following ${\bf Z}_3$ twist:
\begin{equation}
 T_3=(\theta,\theta,\theta \vert\vert \theta,\theta,\theta\vert
 ({1\over 3})^{12} 0^4)~.
\end{equation}   
Here $\theta$ is a $2\pi /3$ rotation of a complex boson (we have complexified 
the original six real bosons into three complex ones). Thus, the first three
entries correspond to the ${\bf Z}_3$ twists of the three right-moving
complex bosons (coming from the six-torus).
The double vertical line separates the right- and left-movers.
The first three left-moving entries correspond to the ${\bf Z}_3$ twists of 
the three left-moving complex bosons (coming from the six-torus). The single 
vertical line separates the latter from the sixteen real bosons corresponding 
to the $\Gamma^{16}$ lattice. The latter are written in the $SO(32)$ basis. 
Thus, for example, $(+1, -1, 0^{14})$ is a root of $SO(32)$ with length 
squared two. There are $480$ roots like this in the $\Gamma^{16}$ lattice, 
and they are descendents of the identity irrep of $SO(32)$. The lattice 
$\Gamma^{16}$ also contains one of the spinor irreps as well. Thus, we will 
choose this spinor irrep to contain the momentum states of the form 
$(\pm {1\over2},...,\pm {1\over 2})$ with even number of plus signs.

{}Note that for the above twist to be a symmetry of the model it is necessary 
(and sufficient) that the twist acting on the $\Gamma^{6,6}$ lattice is a 
rotation in this lattice. This requirement constrains the possible values of 
the metric tensor $g_{ij}$. For the sake of simplicity (this choice will not 
affect how generic the conclusions drawn in this paper will be) let us take the
six-torus to be a product of three two-tori. By $g_{ij}$ we will mean the 
metric of the two-torus then (we can discuss one of the three two-tori, since 
the other two are completely similar). Thus, in this case the metric cannot 
be arbitrary, but must obey the following property: $g_{11}=g_{22}=2R^2$,
$g_{12}=g_{21}=-R^2$, where $R$ is the ``radius'' of the torus. The volume of
the torus, or the K\"ahler structure, is then given by $T=B_{12}+i\sqrt{
\det(g_{ij})}=i\sqrt{3} R^2$ (recall that we have set $B_{ij}=0$). 
The complex structure, however, is fixed: $U=g_{12}/g_{22}+i\det(g_{ij})/g_{22}=
\exp(2\pi i/3)$.

{}Now we are ready to discuss the orbifold model generated by the above twist
$T_3$. This model has $N=1$ space-time supersymmetry, and gauge group
$U(12)\otimes SO(8)$, the same as the type I model discussed in the previous 
section. The untwisted sector gives rise to the $N=1$ supergravity multiplet
coupled to the $N=1$ Yang-Mills gauge multiplet in the adjoint of $U(12)\otimes SO(8)$. 
The matter fields in the untwisted sector are the following chiral 
multiplets: $3({\bf 12}, {\bf 8}_v)(-1)_L$ and $3({\overline {\bf 66}}, {\bf 1})
(+2)_L$. Here the first two entries in the bold font indicate the irreps of the
$SU(12)\otimes SO(8)$ subgroup, whereas the $U(1)$ charges (with the 
normalization radius $1/2\sqrt{3}$) are given in the parenthesis. The factor 
``3'' indicates the number of families (and comes from the right-moving degrees 
of freedom of the heterotic string). There are also chiral multiplets
neutral under the gauge group: $9({\bf 1},{\bf 1})(0)_L$. Note that these contain
18 scalar fields that are the left-over geometric moduli whose vevs parametrize 
the moduli space $SU(3,3,{\bf Z})\backslash SU(3,3)/SU(3)\otimes SU(3)
\otimes U(1)$. (This is the subspace
of the original Narain moduli space $SO(6,6,{\bf Z})\backslash SO(6,6)/SO(6)\otimes 
SO(6)$ that is invariant under the twist.) Actually, the (perturbative) moduli space of this model is larger, and we will return to this point later on.

{}Next, consider the twisted sector. In the twisted sector we have the following 
chiral supermultiplets: $27({\bf 1},{\bf 1})(-4)_L$ and $27({\bf 1},{\bf 8}_s)
(+2)_L$. Here we note that the factor ``27'' comes from the number of fixed points
of the $Z$-orbifold we are considering. 

{}We summarize the massless spectrum of this heterotic string model in Table II.
Note that the $U(1)$ gauge symmetry is anomalous. Thus, the contributions of the untwisted and twisted sectors into the trace anomaly are
$+108$ and $3 \times (+108)$, respectively,
so that the total trace anomaly is $+432$. By the generalized Green-Schwarz mechanism \cite{GS} some of the fields charged under $U(1)$ will acquire vevs to cancel the Fayet-Illiopoulos $D$-term. In the next section we give the superpotential for this model from which it will become clear that the vacuum shift \cite{DSW} can be performed without breaking $N=1$ space-time supersymmetry. (The analysis of the $F$- and $D$-flatness conditions in the type I model was performed in Ref \cite{Sagnotti}. The analysis for the heterotic model is completely analogous.) 

\section{Superpotential}

{}In this section we discuss the perturbative superpotentials for the type I and heterotic string models discussed in the previous sections. Studying the couplings and flat directions in these superpotentials will enable us to make the type I-heterotic duality map more precise.

{}Let us start from the type I model of section II. We refer the reader to Table I for notations. Note that perturbatively the 36 chiral singlets coming from the closed string sector are flat. This can be explicitly seen by computing the scattering amplitudes for these modes within the framework of the conformal field theory of orbifolds. For type IIB such a calculation was performed in Ref \cite{DFMS}. On the other hand, the matter fields coming from the $99$ open string sector have three (and, of course, some higher) point couplings. The lowest order superpotential can be written as
\begin{equation}
 W_I =\lambda \epsilon_{abc} {\mbox{Tr}} (Q_a Q_b \Phi_c)+...~.
\end{equation}
Due to the presence of the anomalous $U(1)$, the Fayet-Illiopoulos $D$-term will have to be cancelled via these charged fields acquiring vevs. This results in breakdown of gauge symmetry, yet the space-time supersymmetry is preserved.

{}Now let us turn to the heterotic string model. The superpotential of this model is more involved than that of the type I model as the twisted sector fields have non-trivial couplings. We will not give the details of calculating these couplings but simply state the results. We refer the reader to the original references \cite{DFMS}, as well as more recent discussion \cite{KST}, where some new tools also were developed. The superpotential for the heterotic string model thus reads (here we are only interested in the general structure of the {\em non-vanishing} terms):
\begin{eqnarray}
 W_H =&&\lambda^\prime  \epsilon_{abc} {\mbox{Tr}}(Q_a Q_b \Phi_c) +\nonumber\\
      && \Lambda_{(\alpha \alpha^\prime \alpha^{\prime\prime})
                   (\beta \beta^\prime \beta^{\prime\prime})
                   (\gamma \gamma^\prime \gamma^{\prime\prime})}
           {\mbox{Tr}} (S_{\alpha \beta \gamma} 
                       T_{\alpha^\prime \beta^\prime \gamma^\prime}
   T_{\alpha^{\prime\prime} \beta^{\prime\prime} \gamma^{\prime\prime}})+...~.
\end{eqnarray}
(The notation can be found in Table II.) Note that the coupling $\Lambda_{(\alpha \alpha^\prime \alpha^{\prime\prime})
(\beta \beta^\prime \beta^{\prime\prime})
(\gamma \gamma^\prime \gamma^{\prime\prime})}
\not=0$ if and only if $\alpha=\alpha^\prime=\alpha^{\prime\prime}$ or $\alpha\not=\alpha^\prime \not=\alpha^{\prime\prime}\not=\alpha$, and similarly for the $\beta$- and $\gamma$-indices. This follows from the orbifold {\em {space group}} selection rules. Here we note that the couplings $\Lambda_{(\alpha \alpha^\prime \alpha^{\prime\prime})
(\beta \beta^\prime \beta^{\prime\prime})
(\gamma \gamma^\prime \gamma^{\prime\prime})}$ with $\alpha\not=\alpha^\prime \not=\alpha^{\prime\prime}\not=\alpha$, and similarly for the $\beta$- and $\gamma$-indices, are exponentially suppressed in the limit of large volume of the compactification manifold, whereas the couplings $\Lambda_{(\alpha \alpha\alpha)(\beta \beta\beta)(\gamma\gamma\gamma)}$ are not. This is because the corresponding $S_{\alpha \beta \gamma}$ and $T_{\alpha \beta \gamma}$ fields are coming from the same fixed point in the latter case, whereas in the former case they are sitting at different fixed points so that upon taking them apart (in the limit of large volume of the orbifold) their coupling becomes weaker and weaker.   

{}Here we can immediately observe that upon the singlets $S_{\alpha \beta \gamma}$, which are the 27 blow-up modes of the $Z$-orbifold (with non-standard embedding) acquiring vevs, the states $T_{\alpha \beta \gamma}$, that transform in the irrep $({\bf 1}, {\bf 8}_s)(+2)$ of $U(12)\otimes SO(8)$, become heavy and decouple from the massless spectrum. Thus, after blowing up the orbifold singularities on the heterotic side (combined with some of the untwisted charged matter fields acquiring vevs to cancel the $D$-term), we can match the massless spectrum to that of the type I model (where the charged matter must acquire vevs to cancel the effect of the anomalous $U(1)$). Note the crucial role of the perturbative superpotential in this matching. It is precisely such that all the extra fields on the heterotic side can be made massive by blowing up the orbifold. 

\section{Moduli Space}

{}Now we turn to the discussion of the moduli spaces for the type I and heterotic models considered in the previous sections. Let us start with the heterotic model. The (perturbative) moduli space of the corresponding Narain model before orbifolding is $SO(6,22,{\bf Z}) \backslash SO(6,22) /SO(6)\otimes SO(22)$. After orbifolding we have two types of moduli: those coming from the untwisted sector, and those coming from the twisted sector. The untwisted sector moduli parametrize the coset $SU(3,9,{\bf Z}) \backslash SU(3,9)/
SU(3) \otimes SU(9) \otimes U(1)$. The subspace $SU(3,3,{\bf Z}) \backslash SU(3,3)/ SU(3) \otimes SU(3) \otimes U(1)$ of this moduli space is parametrized by $18$ neutral singlets $\phi_{ab}$ that correspond to the left-over geometric moduli (coming from the constant metric $g_{ij}$ and antisymmetric tensor $B_{ij}$ fields). The other $36$ moduli correspond to the flat directions in the superpotential for the fields $Q_a$ and $\Phi_a$. (These are the left-over moduli coming from the $6\times 16$ Wilson lines $A^I_i$, $I=1,...,16$, in the
original Narain model).

{}Next, we turn to the twisted moduli of the heterotic string model. First note that in the twisted sector we have the chiral superfields $S_{\alpha\beta\gamma}$ and $T_{\alpha\beta\gamma}$. At the generic point (upon giving appropriate vevs to the $S_{\alpha\beta\gamma}$ fields) the fields $T_{\alpha\beta\gamma}$ become massive (according to the couplings in the superpotential), and there is no superpotential for the $S_{\alpha\beta\gamma}$ fields. There is a subtlety here, however. Since the fields $S_{\alpha\beta\gamma}$ are charged under $U(1)$, a linear combination of them will be eaten in the super-Higgs mechanism. So, naively, not all the $27$ chiral fields survive the Higgsing process. This would pose a problem for matching of the type I and heterotic models. Note, however, that the type I model has anomalous $U(1)$, and to cancel the $D$-term one needs to give vevs to the corresponding charged fields. So, generically, the fields $Q_a$ and $\Phi_a$ will acquire vevs (on the type I side) to break the anomalous $U(1)$. Thus, to match the type I and heterotic models we have to give vevs to the $Q_a$ and $\Phi_a$ fields on the heterotic side as well. Then, we will have $27$ neutral chiral superfields (which, after Higgsing, are a mixture of the original fields $S_{\alpha\beta\gamma}$ and the untwisted matter fields) on the heterotic side, which do correspond to the $27$ neutral chiral superfields coming from the twisted closed string sector of the type I model. Thus, the matching is complete after giving appropriate vevs to {\em both} twisted {\em and} untwisted fields on the heterotic side, as well as giving appropriate vevs to open string sector matter fields, and $27$ twisted closed string moduli. In this process, to make the matching precise, one generically has to appropriately tune the dilaton plus $\phi_{ab}$ geometric moduli on both sides.

{}Thus, the moduli spaces (at generic points) of both type I and heterotic models are the same (at least at the tree-level). They are described by the untwisted moduli of the heterotic string, or equivalently, the moduli coming from the untwisted closed string sector and the open string sector of the type I model (these parametrize the coset $SU(3,9,{\bf Z}) \backslash SU(3,9)/
SU(3) \otimes SU(9) \otimes U(1)$), plus the $2\times 27$ twisted moduli in the heterotic string model (here one needs to take into account the mixing after Higgsing with the untwisted moduli described above), or equivalently, the moduli coming from the twisted closed string sector of the type I model. 
The (perturbative) moduli space (of the heterotic model) is schematically depicted in Fig.1.

{}Once again, we emphasize that the fields $T_{\alpha\beta\gamma}$ are heavy (after appropriate Higgsing), and this is due to the presence of the {\em perturbative} superpotential in the heterotic model. This makes it clear that for $N=1$ type I-heterotic duality the perturbative superpotential is crucial. Another important feature is the presence of anomalous $U(1)$ in these models.
In the type I model the anomalous $U(1)$ forces the open string sector matter fields to acquire vevs to cancel the $D$-term. In the heterotic model we also need to give vevs to the corresponding (untwisted) matter fields in oder to make the matching precise. That is, if we only gave vevs to the charged singlets $S_{\alpha\beta\gamma}$, due to the super-Higgs mechanism some of the degrees of freedom would be eaten in the process of breaking of the anomalous $U(1)$ gauge symmetry. Thus, the matching is only precise if we also involve the untwisted matter fields (on the heterotic side) in cancelling the $D$-term.

{}Note that the appearance of massless twisted matter fields $T_{\alpha\beta\gamma}$ on the heterotic side is a perturbative effect. On the type I side this effect is non-perturbative, and reflects the fact that from type I point of view there is a (non-perturbative) singularity (or, more precisely, a subspace of the full moduli space which is singular) in the moduli space.
Similarly, if we turn on appropriate value for the $B_{ij}$ field on the heterotic side and tune the compactification radii, we can get a {\em perturbative} enhancement of gauge symmetry. Thus, on the heterotic side we can get additional gauge symmetry coming from the six-torus, say, $U(1)^6$. (Actually, this additional gauge symmetry can be as large as $SU(3)^3$ if we start from the $E_6$ six-torus.) On the type I side this is a non-perturbative phenomenon \cite{typeI-het-10}. It requires turning on appropriate constant R-R field $B^\prime_{ij}$.

\section{Discrete Moduli}

{}In this section we discuss the issue of discrete moduli in type I string theory, and their heterotic counterparts. In particular, so far our discussion has been focused on cases where we do not turn on the antisymmetric tensor background on the type I side. Although there are no massless scalars corresponding to these in type I theory (recall that there $B_{ij}$ fields are projected out of the spectrum after orientifolding), {\em i.e.}, these moduli cannot be varied continuously, they can have certain quantized values (because of this they are not moduli in the conventional sense of this word). The quantization is due to the fact that to have a consistent orientifold the corresponding type IIB spectrum must be left-right symmetric. At generic values of $B_{ij}$ this symmetry is destroyed. There are, however, certain discrete $B_{ij}$ backgrounds compatible with the orientifold projection \cite{Sagnotti1}.  

{}Non-zero values of $B_{ij}$ start playing an important role already in the toroidal compactifications of Type I string theory. This issue was addressed some years ago in Ref \cite{Sagnotti1}. If we compactify type I strings on a $d$-torus with zero $B_{ij}$ background, the rank of the gauge group is $d+16$, same as in the heterotic string theory. Here rank 16 comes from the $SO(32)$ gauge group, which might be broken by Wilson lines to its subgroups, whereas rank $d$ comes from the $U(1)^d$ gauge symmetry present in the closed string sector after orientifolding. Upon turning on discrete $B_{ij}$ background, the $U(1)^d$ gauge symmetry is unaffected, yet the rank of the gauge group coming from the $SO(32)$ ({\em i.e.}, Chan-Paton) factor is reduced, depending on the rank $r$ (which is always even, and can take values $2,4,...,d$ for even $d$, and $2,4,...,d-1$ for odd $d$) of the matrix $B_{ij}$. In particular, the rank of the Chan-Paton gauge group is given by $16/2^{r/2}$. 

{}Naturally, one may ask if toroidal type I compactifications have heterotic duals. The answer is quite simple (although constructing dual pairs in certain cases can be non-trivial). The heterotic duals are CHL strings (heterotic strings with maximal supersymmetry) \cite{CHL,CP}. Note that turning on the $B_{ij}$ background does not break supersymmetry (so, for example, a toroidal type I model with $B_{ij}\not=0$ has $N=4$ space-time supersymmetry). Yet the rank of the gauge group is reduced. On the heterotic side this corresponds to turning on {\em discrete} Wilson lines such that at least two of them do not commute with each other (see example below). This is an example of non-Abelian space group orbifold (which is freely acting since its action on the right-movers must be a lattice shift or else fermions would be twisted and the supersymmetry would be reduced). Here we should comment that this is the case if we start from a toroidal compactification of the ${\mbox{Spin}}(32)/{\bf Z}_2$ heterotic string. Then to reduce the rank we need at least two Wilson lines such that they do not commute with each other. Each of these wilson lines by itself does not reduce the rank, and, therefore, can be written in terms of lattice shifts. The point, however, is that they cannot be simultaneously diagonalized in the same basis ({\em i.e.}, they do not commute with each other), hence rank reduction. As we already mentioned, the orbifold in this case is non-Abelian, more precisely, the orbifold space group is non-Abelian. (We refer the reader to the original literature on this subject \cite{NA}, as well as to more recent discussion \cite{kst}, where rules for non-Abelian orbifold model building are derived.)     
Another comment we should make is that if we start from a toroidal compactification of $E_8 \otimes E_8$ heterotic string, the rank reduction without breaking supersymmetry can be achieved by a freely acting orbifold that includes an outer automorphism of the two $E_8$ factors. (If we chose to reinterpret this orbifold in the ${\mbox{Spin}}(32)/{\bf Z}_2$ language, we would ultimately have to consider an orbifold with a non-Abelian space group.) In this case the type I description is no longer adequate, and we must turn to the Type I$^\prime$ description
where part of $E_8 \otimes E_8$ gauge symmetry is non-perturbative and appears due to $D0$-brane dynamics. Note that on the heterotic side we are still having a completely perturbative description in this case.

{}We will not go through construction of dual pairs of maximally supersymmetric type I and heterotic strings in detail except for a brief discussion of one example considered in Ref \cite{Park}. First, let us consider type IIB
theory compactified on $(S^1)^6$. Consider a model obtained from this one by the following freely acting orbifold. Let $\eta_8$ and $\eta_9$ be ${\bf Z}_2$ ({\em i.e.}, order two) shifts in the 8th and 9th directions. Let the ${\bf Z}_2$ orbifold group be ${\cal G}=\{1, \eta_8\eta_9 \}$. The resulting model is a type IIB superstring compactified on $(S^1)^4 \otimes T^2$, with non-zero $B_{89}$ background. Now let us consider the type I orientifold of this model.
The resulting type I string has $U(1)^6 \otimes Sp(16)$ gauge symmetry. The $U(1)^6$ factor comes from the closed string sector, whereas $Sp(16)$ is the Chan-Paton gauge group coming from the open string sector. (In our notations $Sp(2N)$ has rank $N$.) Note that the rank of the $B_{ij}$ matrix in this case is 2, and this model was originally constructed in Ref \cite{Sagnotti1}. There it was also pointed out that by turning on appropriate Wilson lines we can break the Chan-Paton gauge group to $U(16)$ (and in effective field theory language this simply corresponds to the adjoint breaking). If the Wilson line is tuned so that it is ${\bf Z}_2$ valued, the gauge symmetry enhances to $SO(16)$. In this limit we recover the model of Ref \cite{Park}, where this model was constructed by starting from type IIA superstring, twisting by the above orbifold ${\cal G}$, and then orientifolding. In this case we have a type I$^\prime$ description. 

{}Constructing the heterotic dual of the above type I model with $SO(16)$ gauge symmetry is not difficult. Thus, consider heterotic string compactified on $(S^1)^6$. Let us turn on the following two Wilson lines:
\begin{eqnarray}
 &&U_1=(0^4,-e_5/2,0 \vert\vert 0^4, e_5/2,0\vert
 ({1\over 4} ,-{1\over 4})^8)~,\\
 &&U_2=(0^4,0,-e_6/2 \vert\vert 0^4,0,e_6/2\vert
 \sigma^8)~,
\end{eqnarray}
where the coordinates corresponding to the compactification six-torus are written in real basis, and $e_i$ are defined as 
in Eq (\ref{momenta}). (Here the lattice is cubic, {\em i.e}, $e_i\cdot e_j =0$
for $i\not=j$.) Thus, the last two of the coordinates of $(S^1)^6$ are shifted by ${\bf Z}_2$ shifts ($(-e_5/2\vert\vert e_5/2)$ for the fifth $S^1$, and $(-e_6/2\vert\vert e_6/2)$ for the sixth $S^1$).
As far as the ${\mbox{Spin}}(32)/{\bf Z}_2$ lattice is concerned, the $U_1$ Wilson line shifts it by a ${\bf Z}_2$ shift (recall that $(({1\over 2} ,-{1\over 2})^8)$ is one of the spinorial weights of ${\mbox{Spin}}(32)/{\bf Z}_2$). On the other hand, the action of $U_2$ is specified by the action of the ${\bf Z}_2$ {\em twist} $\sigma=\left( \begin{array}{cc}
0& 1\\1& 0               \end{array}
\right)$ that interchanges the corresponding two real chiral world-sheet bosons. Note that $\sigma^2={\mbox{diag}}(1,1)$, so that $U_2$ is a ${\bf Z}_2$ Wilson line as well. Also note that 
just the action of $U_1$ alone breaks $SO(32)$ to $U(16)$. Similarly, just the action of $U_1$ alone breaks $SO(32)$ to $U(16)$ (when acting alone, $U_2$ can be diagonalized, and its action on the ${\mbox{Spin}}(32)/{\bf Z}_2$ lattice is the same as that of $U_1$). The crucial point, however, is that $U_1$ and $U_2$ Wilson lines cannot be diagonalized simultaneously, {\em i.e.}, they (or, more precisely, their actions on the ${\mbox{Spin}}(32)/{\bf Z}_2$ lattice) do not commute. Nonetheless, this is consistent since they do not commute up to the shift $(({1\over 2}, -{1\over 2})^8)$ which belongs to the ${\mbox{Spin}}(32)/{\bf Z}_2$ lattice. The action of the $U_1$ and $U_2$ Wilson lines on the ${\mbox{Spin}}(32)/{\bf Z}_2$ lattice is thus isomorphic to that of $D_4$ subgroup of $SU(2)$. ($D_4$ is generated by two elements $p$ and $q$ such that $p^4=q^2=1$, and $pq=qp^3$. In our case $q$ is $\sigma$, and $p$ is the $({1\over 4}, -{1\over 4})$ shift. Note that $p^4$ belongs to the root lattice of $SO(32)$, but $p^2$ is actually already in the ${\mbox{Spin}}(32)/{\bf Z}_2$ lattice since $p^2$ is a spinorial weight of the latter.) The fact that $U_1$ and $U_2$ do not commute leads to rank reduction for the gauge group. The reader can easily work out the final gauge group to be $SO(16)$. (This $SO(16)$ is realized via level-2 current algebra on the world-sheet. See Ref \cite{kst} for details.) Note that the action of $U_1$ and $U_2$ on $(S^1)^6$ is that of ${\bf Z}_2 \otimes {\bf Z}_2$, not of $D_4$. This is possible because a ${\bf Z}_4$ shift in the root lattice of $SO(32)$ can be a ${\bf Z}_2$ shift in the ${\mbox{Spin}}(32)/{\bf Z}_2$ lattice for the latter contains the spinorial weights on top of the root weights of $SO(32)$. (Here the situation is analogous to that in the ${\bf Z}_2\otimes {\bf Z}_2$ type I model of Ref \cite{BL}. In that model the action of the orbifold on the compactified coordinates is that of ${\bf Z}_2\otimes {\bf Z}_2$, whereas the action on the Chan-Paton factors is that of $D_4$. Instead of the breaking $SO(32)\supset U(16)\supset SO(16)$ as in the above heterotic model, in the type I model we get $Sp(16)$ subgroup via the breaking $SO(32)\supset U(16)\supset Sp(16)$.)

{}Finally, we comment on $N=1$ type I models with reduced rank. (Type I compactifications on $K3$ with non-trivial antisymmetric tensor field flux have been recently discussed 
in Ref \cite{sethi}.) It is clear from the above discussion, that compactifying a type I model on an orbifold with a non-zero discrete $B_{ij}$ background should be dual to a heterotic string model on an orbifold with a non-Abelian space group. One must be careful, however, with precise duality matchings in this case as models (on both sides) get more and more complicated. Here we give an example of a heterotic string model with reduced rank that does {\em not} correspond to a type I model with non-zero $B_{ij}$ background. Thus, let us start from the Narain model of section III, and orbifold it by the following twist: 
\begin{equation}
 T^\prime_3=(\theta,\theta,\theta \vert\vert \theta,\theta,\theta\vert
 {\cal P}\vert({1\over 3})^4)~.
\end{equation}   
Here we are writing the ${\mbox{Spin}}(32)/{\bf Z}_2$ momenta in the $SO(32)\supset SO(8)^4$ basis. Then ${\cal P}$ is the twist that acts as an outer automorphism on the first three $SO(8)$ subgroups, whereas $({1\over 3})^4$ is a shift in the last $SO(8)$ subgroup. This model is the same as that of section III. Now add the following Wilson lines:   
\begin{eqnarray}
 &&U_1=(-e_1/2,0,0 \vert\vert e_1/2,0,0\vert
 ({1\over 2})^4,0^4,0^4\vert (-{1\over 2})^4)~,\\
 &&U_2=(-e_2/2,0,0 \vert\vert e_2/2,0,0\vert
 0^4,({1\over 2})^4,0^4\vert (-{1\over 2})^4)~,
\end{eqnarray}  
Note that $U_1$ and $U_2$ generate a ${\bf Z}_2 \otimes {\bf Z}_2$ orbifold (for $U_1$and $U_2$ are order two shifts commuting with each other).
It is easy to see that these Wilson lines respect the original ${\bf Z}_3$ symmetry of the orbifold (note, in particular, that $\theta U_1=U_2$ and $\theta U_2=-U_1-U_2$), and all the other string consistency conditions 
(such as level matching, for example) are satisfied as well. The final orbifold group generated by the elements $T^\prime_3$, $U_1$ and $U_2$ is the tetrahedral group, which is a non-Abelian finite discrete subgroup (from the E-series in the A-D-E classification) of $SU(2)$. The resulting model has $N=1$ supersymmetry and $SO(8) \otimes U(4)$ gauge group. Note that the $SO(8)$ factor is realized via the corresponding Kac-Moody algebra at level 3 (note that similar structures have recently appeared in construction of three-family grand unified string theories \cite{kt}), whereas $SU(4)$ is a level-1 subgroup. The massless spectrum of this model is given in Table III (where the $U(1)$ normalization radius is $1/6$). Note that this model has reduced rank and naively one might suspect that its type I dual is a model with non-zero $B_{ij}$ background, but this is not so. This heterotic string model belongs to the same moduli space as the model discussed in section III, {\em i.e.}, the moduli space ${\cal M}$ in Fig.1. More concretely, it belongs to the subspace $C\subset {\cal M}$. If we give the singlets $S_{\alpha\beta\gamma}$ appropriate vevs we can move to the subspace $A$, where the dual type I model sits. This model can be obtained from the type I model discussed in section II by turning on the appropriate vevs for the fields $\Phi_a$ and $Q_a$. Thus, the rank reduction on the heterotic side can often be achieved within the same moduli space. The type I moduli spaces with different $B_{ij}$, however, seem to be disjoint, just like those of CHL heterotic strings.

{}It would be interesting and instructive to understand if there exist (modular invariant) heterotic duals of the reduced rank $N=1$ type I models of Ref \cite{Sagnotti}. In particular, there is a model with $SO(8) \otimes U(4)$ gauge group (same as of the above heterotic string model) with three generations of chiral multiplets (coming from the 99 open string sector) in irreps $({\bf 8}_v,{\bf 4})_L$ and $({\bf 1},{\overline {\bf 10}})_L$ (here we drop the $U(1)$ charges). 

\section{Type I-Heterotic Dictionary}

{}In this section we summarize the observations discussed in the previous sections into a dictionary that translates type I effects into heterotic ones, and vice-versa. The dictionary is given in Table IV. Here we would like to make some comments. First of them concerns the last line in Table IV, {\em i.e.}, the relation between small instantons on the heterotic side and $D5$-branes on the type I side. The statement that $D5$-branes are perturbative from type I point of view is precise. However, whether they correspond to non-perturbative effects on the heterotic side deserves some explanation. The 55 open string states do correspond to non-perturbative effects on the heterotic side. The 59 strings, however, are perturbative from the heterotic point of view in the sense that the 59 matter (after Higgsing) is present in the {\em twisted} sectors of the corresponding heterotic dual (and type I superstring compactified on the ${\bf Z}_2$ orbifold limit of $K3$ \cite{GP} is an example of this \cite{6}). 
We would also like to point out that enhancement of gauge symmetry by tuning $B_H$ (and the radii) on the heterotic side is perturbative, whereas the corresponding enhancement on the type I side (which requires turning on the appropriate $B^\prime_I$ background) is not. Finally, we mention that the 99 open string states correspond to the untwisted states of the heterotic dual (whenever the latter exists), except for the geometric moduli of the heterotic model whose duals are found in the closed string sector of the type I model.  

\section{Conclusions}

{}In this paper we have considered some aspects of $N=1$ type I-heterotic duality in four dimensions. We have confined our attention on effects that are perturbative on the heterotic side, and can be perturbative or non-perturbative on the type I side. This was done with the goal in mind that once we understand the relation between such effects in the two descriptions, we might be able to learn about the non-perturbative effects in the heterotic string theory by mapping them onto perturbative effects (such as $D5$-branes) on the type I side. One of the steps that can be made in these directions is to construct an $N=1$ type I model with both $D9$- and $D5$-branes, separate the effects that are perturbative on the heterotic side from those that are not, and attempt to construct the heterotic string model (being guided by perturbative part) whose extension into the regions in the moduli space where the perturbation theory breaks down would be the candidate heterotic dual of the type I model. Then we can try to learn about the non-perturbative heterotic string effects by comparing them with their perturbative type I counterparts.

{}We can see at least one way of making this possible. Consider type I theory on (there are two inequivalent possibilities here) a (symmetric) ${\bf Z}_6$ orbifold that breaks supersymmetry to $N=1$. The corresponding open string sector will have both $D9$- and $D5$-branes. Constructing (modular invariant) candidate heterotic dual in this case should not be difficult (in the region where the perturbation theory is good). Then one could try to go into the non-perturbative regime by tuning certain moduli. The type I model should be a rather good guiding principle (provided that the type I-heterotic duality still holds, at least in some sense). Here we note that the ${\bf Z}_2 \otimes {\bf Z}_2$ orbifold model (which has $N=1$ SUSY and $Sp(2N)$ gauge groups) of Ref \cite{BL} does not appear to have a modular invariant heterotic dual.   

{}We should mention that in this paper we only considered tree-level effects. In particular, the matching of the spectrum and moduli spaces has been done only at tree-level. It would be important to understand (perturbative) quantum corrections (and their matching, if any) to, say, K{\"a}hler potential. This should be within our present technology.  Note that the quantum corrections in the $N=2$ case have been studied in detail in Ref \cite{taylor}.  

\acknowledgements

{}We would like to thank Damiano Anselmi, Ignatios Antoniadis, Costas Bachas, Michael Bershadsky, Robbert Dijkgraaf, Gia Dvali, Andrei Johansen, Luis Ib{\'a}{\~n}ez, Costas Kounnas, Augusto Sagnotti, Tom Taylor, Henry Tye, Angel Uranga, Cumrun Vafa, Erik Verlinde and Herman Verlinde for discussions. The work of Z.K. was supported in part by the grant NSF PHY-96-02074, and the DOE 1994 OJI award. Z.K. would like to thank the Institute for Theoretical Physics at the University of Amsterdam, and the Theoretical Physics Division at CERN for their kind hospitality while parts of this work were completed.  Z.K. would also like to thank Albert and Ribena Yu for financial support.

\begin{table}[t]
\begin{tabular}{|c|c|l|l|}
Sector & Field & $SU(12)\otimes SO(8) \otimes U(1)$ & Comments\\
\hline
Closed & & &\\
Untwisted & $\phi_{ab}$ & $9({\bf 1}, {\bf 1})(0)_L$ & $a,b=1,2,3$\\
\hline
Closed & & &\\
Twisted & $S_{\alpha\beta\gamma}$ & $27({\bf 1}, {\bf 1})(0)_L$ & $\alpha,\beta,\gamma=1,2,3$ \\
\hline
Open & $Q_a$ & $3({\bf 12},{\bf 8}_v)(-1)_L$ & \\
     & $\Phi_a$ & $3({\overline {\bf 66}}, {\bf 1})(+2)_L$ & $a=1,2,3$\\
\end{tabular}
\caption{The massless spectrum of the type I model with $N=1$ space-time supersymmetry and gauge group $SU(12)\otimes SO(8) \otimes U(1)$ discussed in section II. The gravity, dilaton and gauge supermultiplets are not shown.}  
\end{table}

\begin{table}[t]
\begin{tabular}{|c|c|l|l|}
Sector & Field & $SU(12)\otimes SO(8) \otimes U(1)$ & Comments\\
\hline
  & $\phi_{ab}$ & $9({\bf 1}, {\bf 1})(0)_L$ & $a,b=1,2,3$\\
Untwisted & $Q_a$ & $3({\bf 12},{\bf 8}_v)(-1)_L$ & \\
     & $\Phi_a$ & $3({\overline {\bf 66}}, {\bf 1})(+2)_L$ & \\
\hline
Twisted & $S_{\alpha\beta\gamma}$ & $27({\bf 1}, {\bf 1})(-4)_L$ & $\alpha,\beta,\gamma=1,2,3$ \\
   & $T_{\alpha\beta\gamma}$ & $27({\bf 1}, {\bf 8}_s)(+2)_L$ & \\
\end{tabular}
\caption{The massless spectrum of the heterotic model with $N=1$ space-time supersymmetry and gauge group $SU(12)\otimes SO(8) \otimes U(1)$ discussed in section III. The gravity, dilaton and gauge supermultiplets are not shown.}  
\end{table}

\newpage

\begin{table}[t]
\begin{tabular}{|c|c|l|l|}
Sector & Field & $SO(8)\otimes SU(4) \otimes U(1)$ & Comments\\
\hline
  & $\phi_{ab}$ & $9({\bf 1}, {\bf 1})(0)_L$ & $a,b=1,2,3$\\
Untwisted & $Q_a$ & $3({\bf 1},{\bf 6})(+2)_L$ & \\
     & $\Phi_a$ & $3({{\bf 28}}, {\bf 1})(0)_L$ & \\
\hline
Twisted & $S_{\alpha\beta\gamma}$ & $27({\bf 1}, {\bf 1})(-4)_L$ & $\alpha,\beta,\gamma=1,2,3$ \\
   & $T_{\alpha\beta\gamma}$ & $27({\bf 8}_s,{\bf 1})(+2)_L$ & \\
\end{tabular}
\caption{The massless spectrum of the heterotic model with $N=1$ space-time supersymmetry and gauge group $SO(8)\otimes SU(4) \otimes U(1)$ discussed in section VI. The gravity, dilaton and gauge supermultiplets are not shown.}  
\end{table}

\begin{table}[t]
\begin{tabular}{|l|l|l|l|}
Type I & Heterotic & Origin & Comments\\
\hline
$\phi_I$ & $\phi_H$ & P/P &
 $\phi_H={1\over 2}\phi_I -{1\over 8}\log(\det(g_I))$ \\
\hline
$g_I$ & $g_H$ & P/P & \\
\hline
$B^\prime_I$ & $B_H$ &NP$^\ast$/P & $^\ast$see comments in section VII\\
\hline
$A_I$ & $A_H$ & P/P & Wilson lines\\
\hline
$B_I$ & ${\tilde A}_H$ & P/P & Discrete Wilson lines in heterotic\\
\hline
$D9$-branes & $SO(32)$ & P/P & Chan-Paton gauge group in type I\\
\hline 
$D5$-branes & small instantons & P/NP$^\ast$ & $^\ast$see comments in section VII\\
\end{tabular}
\caption{The dictionary between type I and heterotic superstrings. P and NP stand for perturbative and non-perturbative, respectively. Thus, NP/P means non-perturbative in type I, and perturbative in heterotic.}  
\end{table}

\newpage
\begin{figure}[t]
\epsfxsize=16 cm
\epsfbox{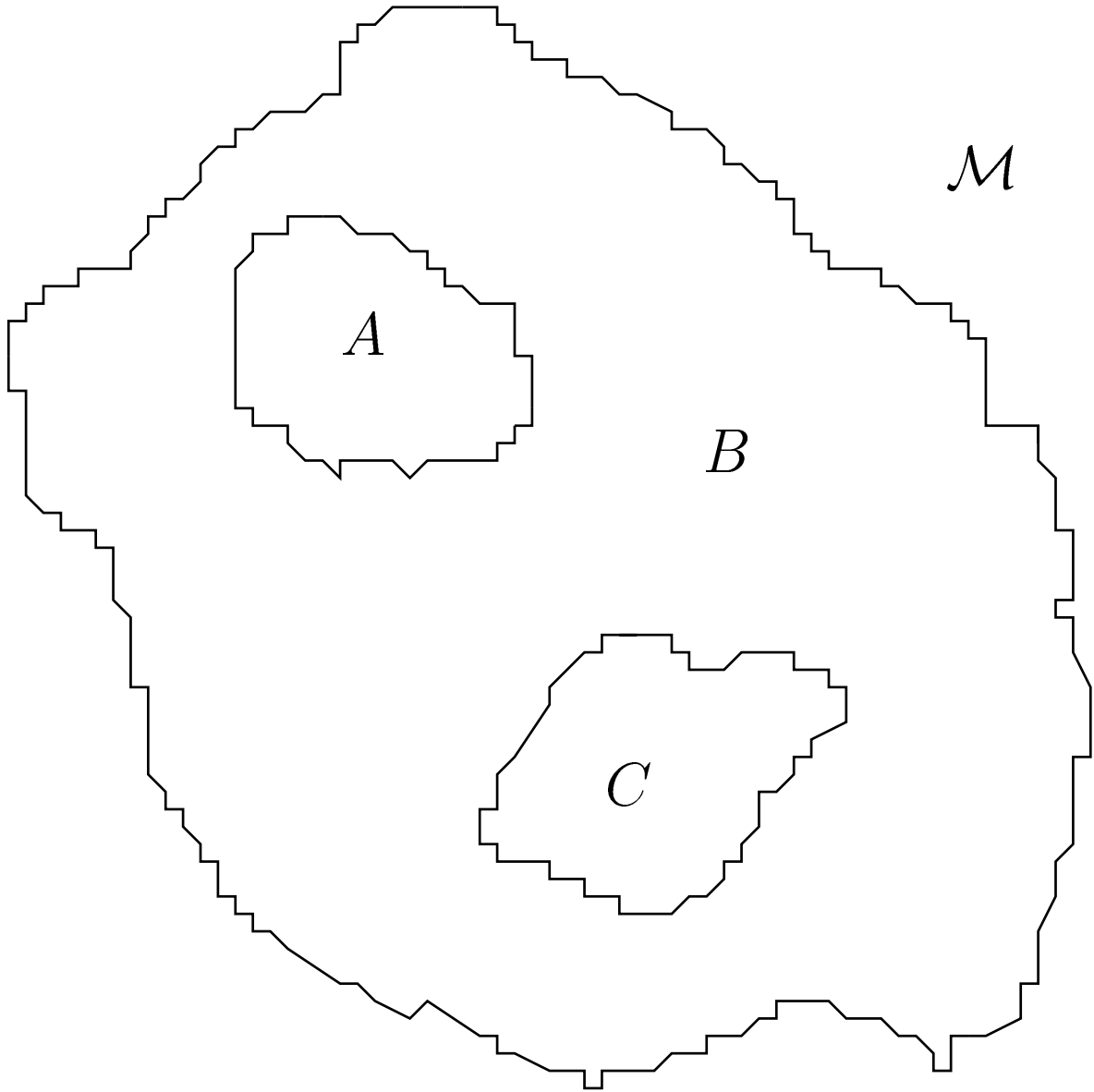}
\caption{A schematic picture of the (perturbative) moduli space ${\cal M}$
(of the heterotic model). Region $A$ is the subspace corresponding to the type I model. Region $C$ is the subspace where some or all of the $S_{\alpha\beta\gamma}$ vevs are zero and some or all of the $T_{\alpha\beta\gamma}$ fields are massless. Region $B$ complements $A$ and $C$ in ${\cal M}$.}
\end{figure}

\end{document}